\documentclass[aps,pra,reprint,superscriptaddress,floatfix]{revtex4-2}
\usepackage{graphicx}
\usepackage{dcolumn}
\usepackage{gensymb} 
\usepackage{ulem}
\usepackage{amsmath,amsfonts,amssymb,bm}
\usepackage{graphics}			
\usepackage{xcolor}				
\usepackage[caption=false]{subfig}

\newcommand{\rhs}{Rh$_{17}$S$_{15}$}

\begin{document}

\title{Unconventional nodal superconductivity in miassite Rh$_{17}$S$_{15}$}

\author{Hyunsoo Kim}
\altaffiliation{Present affiliation: Department of Physics, Missouri University of Science and Technology, Rolla, Missouri 65409, USA}
\affiliation{Ames National Laboratory, Iowa State University, Ames, Iowa 50011, USA}
\affiliation{Department of Physics \& Astronomy, Iowa State University, Ames, Iowa 50011, USA}

\author{Makariy A. Tanatar}
\affiliation{Ames National Laboratory, Iowa State University, Ames, Iowa 50011, USA}
\affiliation{Department of Physics \& Astronomy, Iowa State University, Ames, Iowa 50011, USA}

\author{Marcin Ko{\'n}czykowski}
\affiliation{Laboratoire des Solides Irradi{\'e}s, CNRS UMR 7642 \& CEA-DSM-IRAMIS, {\'E}cole Polytechnique, F-91128 Palaiseau cedex, France}

\author{Udhara S. Kaluarachchi}
\affiliation{Ames National Laboratory, Iowa State University, Ames, Iowa 50011, USA}
\affiliation{Department of Physics \& Astronomy, Iowa State University, Ames, Iowa 50011, USA}

\author{Serafim Teknowijoyo}
\affiliation{Ames National Laboratory, Iowa State University, Ames, Iowa 50011, USA}
\affiliation{Department of Physics \& Astronomy, Iowa State University, Ames, Iowa 50011, USA}

\author{Kyuil Cho}
\affiliation{Ames National Laboratory, Iowa State University, Ames, Iowa 50011, USA}

\author{Aashish Sapkota}
\affiliation{Ames National Laboratory, Iowa State University, Ames, Iowa 50011, USA}

\author{John M. Wilde}
\affiliation{Ames National Laboratory, Iowa State University, Ames, Iowa 50011, USA}
\affiliation{Department of Physics \& Astronomy, Iowa State University, Ames, Iowa 50011, USA}

\author{Matthew J. Krogstad}
\affiliation{Materials Science Division, Argonne National Laboratory, Lemont, Illinois 60439, USA }

\author{Sergey L. Bud'ko}
\affiliation{Ames National Laboratory, Iowa State University, Ames, Iowa 50011, USA}
\affiliation{Department of Physics \& Astronomy, Iowa State University, Ames, Iowa 50011, USA}

\author{Philip M. R. Brydon}
\affiliation{Department of Physics and MacDiarmid Institute for Advanced Materials and Nanotechnology, University of Otago, P.O. Box 56, Dunedin 9054, New Zealand}

\author{Paul C. Canfield}
\affiliation{Ames National Laboratory, Iowa State University, Ames, Iowa 50011, USA}
\affiliation{Department of Physics \& Astronomy, Iowa State University, Ames, Iowa 50011, USA}

\author{Ruslan Prozorov}
\email[corresponding author: ]{prozorov@ameslab.gov}
\affiliation{Ames National Laboratory, Iowa State University, Ames, Iowa 50011, USA}
\affiliation{Department of Physics \& Astronomy, Iowa State University, Ames, Iowa 50011, USA}

\date{\today}

\begin{abstract}
Unconventional superconductivity has long been believed to arise from a lab-grown correlated electronic system. Here we report compelling evidence of unconventional nodal superconductivity in a mineral superconductor \rhs. We investigated the temperature-dependent London penetration depth $\Delta\lambda(T)$ and disorder evolution of the critical temperature $T_c$ and upper critical field $H_{c2}(T)$ in synthetic miassite \rhs.
We found a power-law behavior of $\Delta\lambda(T)\sim T^n$ with $n\approx 1.1$ at low temperatures below $0.3T_c$ ($T_c$ = 5.4 K), which is consistent with the presence of lines of the node in the superconducting gap of \rhs.
The nodal character of the superconducting state in \rhs~was supported by the observed pairbreaking effect in $T_c$ and $H_{c2}(T)$ in samples with the controlled disorder that was introduced by low-temperature electron irradiation.
We propose a nodal sign-changing superconducting gap in the $A_{1g}$ irreducible representation, which preserves the cubic symmetry of the crystal and is in excellent agreement with the superfluid density, $\lambda^2(0)/\lambda^2(T)$.
\end{abstract}


\maketitle

Superconductivity is extremely rare in naturally occurring compounds. 
{Elemental metals are prone to producing oxides, sulfates, nitrates, etc, while elemental lead and tin exist in single crystalline form in some ores as rare exceptions.}
Their mixture with indium is responsible for traces of superconductivity found in meteorites \cite{meteorite}. 
{The two minerals known to show superconductivity are covellite, CuS, with superconducting transition temperature $T_c=$1.6~K \cite{covellite,DI-BENEDETTO2006} and parkerite, Ni$_3$Bi$_2$S$_2$, with $T_c\approx$0.7~K \cite{Sakamoto2006,Lin2012}.} 
Another naturally occurring mineral, miassite, is a rhodium sulfide, and one of the very few rhodium minerals known. 
This compound was identified in the 30's \cite{RhSphasediagram}, and superconductivity in polycrystals, of then believed at the time to have the Rh$_9$S$_8$ composition, was reported in 1954 by Matthias {\it et al.} \cite{Matthias1954}. 
It was one of the first sulfide superconductors ($T_c$=5.8~K) known. 
The composition was refined to Rh$_{17}$S$_{15}$ in the later crystal structure description \cite{Geller1962}.  
The mineral with the same composition was discovered significantly later, in the 80's \cite{mineraldiscovery} in the placers of Miass river in Ural mountains and got its name after it \cite{mineraldiscovery,Naren2008}.  
Natural miassite is found in isoferroplatinum deposits as small, rounded inclusions of up to 100 $\mu$m in diameter.
While superconductivity in tin, lead, and covellite \cite{DI-BENEDETTO2006,Casaca2011} can be suppressed by a low magnetic field, miassite exhibits an anomalously high upper critical field greater than 20 T \cite{Settai2010}, which significantly exceeds the paramagnetic Pauli limit \cite{Naren2008}.
In this report, we study the superconducting gap structure in miassite crystals, analyzing low-temperature London penetration depth, and probing the response of its superconducting transition temperature, $T_c$, and of the upper critical field, $H_{c2}$, to nonmagnetic disorder induced by 2.5 MeV electron irradiation. 
We find that miassite has a nodal superconducting gap, similar to that found in the cuprates, organic, and some heavy fermion superconductors.  
Our finding identifies miassite as so far the only naturally occurring nodal superconductor and suggests that unconventional superconductivity may be a much more generic phenomenon in nature.

Superconductivity is the phenomenon observed in metals, in which the resistivity falls abruptly to zero below a certain critical temperature $T_c$.
The superconducting transition is accompanied by magnetic flux expulsion from the bulk (Meissner effect) of the sample \cite{Meissner1933}.  
Superconductivity results from the instability of the metallic state to Cooper pairing of conduction electrons as explained by Bardeen, Cooper, and Schrieffer (BCS) in 1957 \cite{Bardeen1957}. This instability opens a gap $\Delta$ on the Fermi surface.
The BCS theory has been generalized to multiband systems where there are different uniform gap magnitudes on the different Fermi surfaces, e.g. MgB$_2$  \cite{Bouquet2001, Canfield2003}. However, it cannot explain the sign-changing gap which is a hallmark of ``unconventional" superconductors, which is observed in several classes of superconductors including the high $T_c$ cuprates \cite{Hardy1993} and iron-based compounds \cite{Hicks2009}. 

In this work, we report unconventional superconductivity in the synthetic compound Rh$_{17}$S$_{15}$, which has the same chemical formula as the naturally occurring mineral miassite. 
Our discovery is based on the measurements of the temperature-dependent London penetration depth and the response of superconducting transition temperature to non-magnetic disorder. London penetration depth shows $T-$linear variation below $T_c/3$ establishing the presence of low-energy excitations down to the lowest temperature of $T_c/100$. This is consistent with the line nodes in the superconducting order parameter. Another key evidence of unconventional behavior is observing a significant suppression of $T_c$ by non-magnetic defects (mostly vacancies) induced by 2.5 MeV electron irradiation.
These results clearly establish Rh$_{17}$S$_{15}$ as an unconventional nodal superconductor. 
Our theoretical analysis is consistent with an extended $s$-wave state possessing circular line nodes, but cannot rule out a nematic $d$-wave state.
Rhodium-based superconductors are relatively rare, and the observation of unconventional superconductivity in Rh$_{17}$S$_{15}$ provides new insights into the underlying mechanisms governing this phenomenon.

To investigate the superconducting state in \rhs, we synthesized single crystalline samples out of the Rh-S eutectic region by using a high-temperature flux growth technique. In Ref. \cite{Lin2012}, it has been shown that the high-temperature solution growth technique can be used to grow binary and ternary transition metal-based compounds out of S-based solutions. In Refs. \cite{Kaluarachchi2015,Kaluarachchi2016}, high temperature solution growth was expanded to Rh-S-X ternaries.  As part of that effort, we re-determined the Rh-rich eutectic composition to be close to Rh$_{60}$S$_{40}$.  As a result, we were able to create a slightly more S-rich melt, Rh$_{58}$S$_{42}$, by combining elemental Rh powder (99.9+ purity) and elemental S in a fritted Canfield Crucible set \cite{Canfield2016}, sealing in a silica ampoule, slowly heating (over 12 hours) to 1150\degree C and then slowly cooling from 1150 to 920\degree C over 50 hours and decanting \cite{Canfield2019}.   Millimeter-sized single crystals of \rhs~grew readily (see inset to Fig. 1a). 

London penetration depth measurements were made using tunnel diode resonator (TDR) technique \cite{Van-Degrift1975} in dilution refrigerator to access temperatures significantly below $T_c$, as low as $T_c/100$ \cite{kim2018} (see Appendix for details).
The non-magnetic scattering in Rh$_{17}$S$_{15}$ was introduced by low-temperature electron irradiation controlling the disorder in the samples, and the superconducting $T_c$ is reduced by 26\% and 40\% after irradiation with relativistic electrons with doses of 0.912 C/cm$^2$ and 2.912 C/cm$^2$, respectively. Resistivity measurements were made on the same samples before and after irradiation to exclude uncertainty of the geometric factor determination. 
Analysis of the normalized temperature-dependent superfluid density and the $T_c$ suppression rate suggests that the data are best described by an extended $s$-wave superconducting state with accidental line nodes within a cubic symmetry. 
This is in line with the parallel shift of the $H_{c2}(T)$ curves with the increase of disorder, revealing the pair-breaking character of non-magnetic scattering.


\begin{figure}
\includegraphics[width=0.8\linewidth]{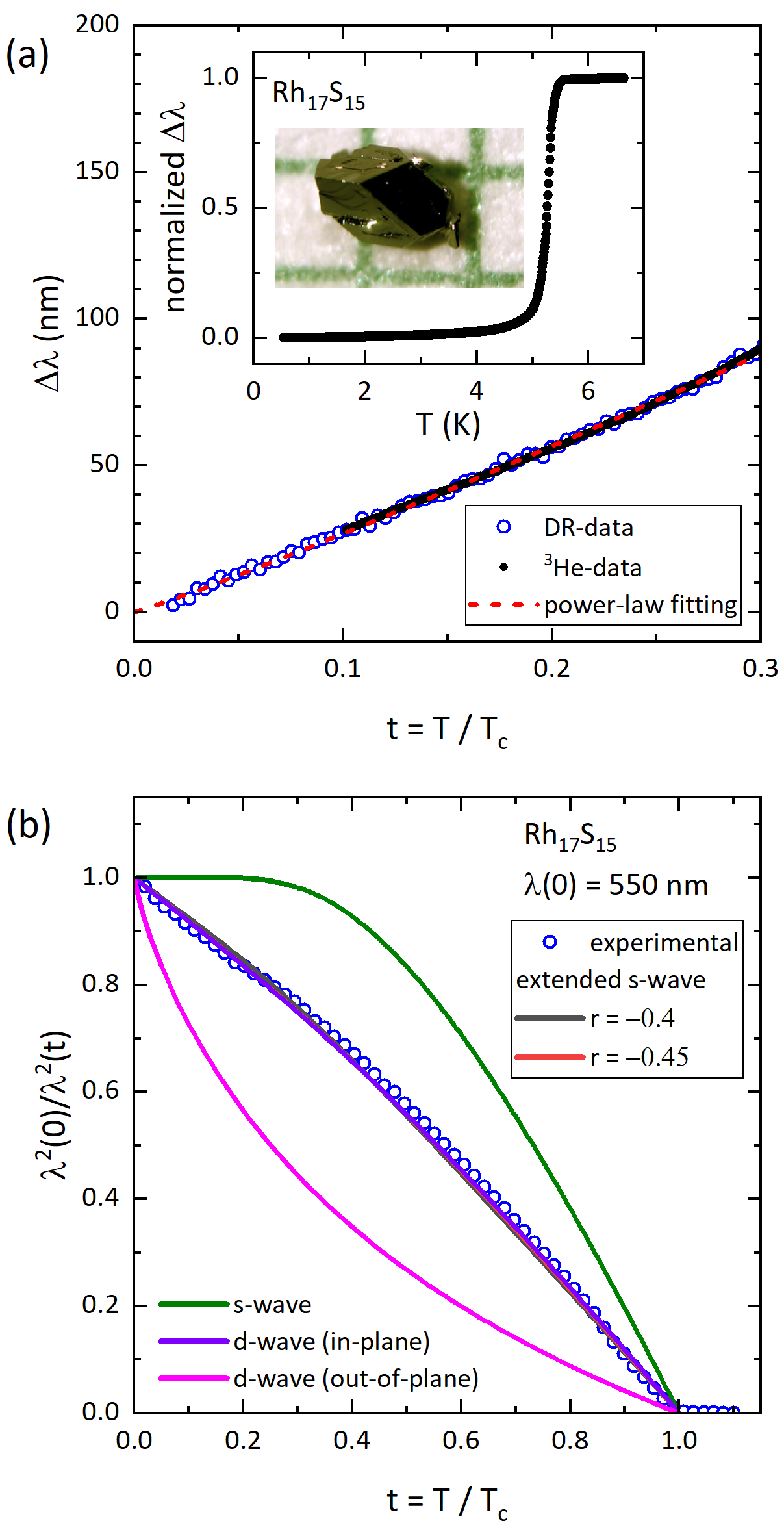}%
\caption{\label{fig1} London penetration depth and superfluid density in \rhs. (a) The temperature variation of London penetration depth $\Delta\lambda(T)$ in an as-grown single crystal \rhs. The main panel shows $\Delta\lambda(T)$ in a characteristic temperature range below $0.3T_c$. The open and closed symbols represent data taken by DR-TDR and $^3$He-TDR, respectively. The dashed line represents the best fit of the power-law function. Inset: the normalized data $\Delta\lambda(T)/\Delta\lambda(6\textmd{K})$ and photo of a typical single crystal.
(b) Normalized superfluid density $\rho_s=\lambda^2(0)/\lambda^2(T)$ in \rhs. The lines represent the theoretical superfluid density curves for full-gap $s$-wave (green), line-nodal $d$-wave in-plane (violet), out-of-plane (magenta), and anisotropic $s$-wave states (grey and red). $r$ is defined in Eq. \ref{eq:gap}. We note that the curves for $d$-wave and extended $s$-wave are nearly identical.
}
\end{figure}

Figure \ref{fig1}a shows 
the London penetration depth $\Delta \lambda(T)$ in an as-grown single crystalline sample (S1). As shown in the inset, the pristine sample exhibits a sharp superconducting transition at $T_c=5.31$ K,  determined by using a maximum of $\Delta\lambda(T)$ derivative criterion. 
In the main panel, we zoom in on the low-temperature behavior of $\Delta \lambda(T)$ at temperatures below $T<0.3T_c$. The sample S1 was measured in both a $^3$He cryostat (black full symbol) and {dilution of refrigerator} \cite{RSI} (blue open symbol). 
{Measurements of $\Delta \lambda(T)$} in this temperature range do not depend on the temperature evolution of the superconducting gap but rather reflect the gap structure.  
Fully gapped superconductors would exhibit an exponential saturation below $T<0.3T_c$, following a relation,
\begin{equation}\label{eq:swave}
   \Delta\lambda(T)=\lambda(0) \sqrt{\frac{\pi \Delta_0}{2k_BT}} \exp{\left(-\frac{\Delta_0}{k_BT}\right)} 
\end{equation}
where $\lambda(0)$ and $\Delta_0$ represents the penetration depth at absolute zero and the maximum gap magnitude, respectively.
Instead, the observed $\Delta\lambda(T)$ curves in \rhs~are very close to $T$-linear which is expected for superconductors with line nodes and found, for example, in the cuprates \cite{Hardy1993}. This nearly $T$-linear behavior extends down to very low temperatures ($\sim T_c/100$) ruling out the existence of deep gap minima and is distinctly different from suggested $s$-wave pairing \cite{Koyama2010}. 
We fit the low-temperature data up to $0.3T_c$ using a power-law function, $\Delta\lambda(T)=AT^n$, with both $n$ and $A$ as free fitting parameters. 
We obtained $n\approx 1.1$ and $A\approx 55$ nm/K$^n$. 
A small deviation from the $T$-linear behavior in a line-nodal superconductor can arise due to the impurity scattering \cite{Hirschfeld1993, Cho2022}.
However, the coherence length $\xi(0)\approx 4$ nm \cite{Settai2010} is much smaller than the mean free path of typical single crystal samples, and therefore our single crystal \rhs~sample is likely in a clean limit.

The superconducting gap structure can be analyzed using the temperature-dependent superfluid density, $\rho_s(T) = \lambda^2(0)/\lambda^2(T)$. The absolute value of $\lambda$ at $T=0$ is not found in our experiment. The reported values of $\lambda(0)$ in the literature vary from 490 nm \cite{Settai2010} to 700 nm \cite{Naren2011}. The former was determined from an experimental $H_{c1}(0)=30$ Oe within the Ginzburg-Landau theory \cite{Settai2010}, and the latter from $\mu$SR studies \cite{Naren2011}.  For the calculation of the superfluid density in  Fig. \ref{fig1}b, we adopted $\lambda(0)=550$ nm (open symbols). This value is the most compatible with the thermodynamic quantities in the Rutgers relation  $\rho'_s(1)/\lambda^2(0)=16\pi^2 T_c \Delta C/\phi_0 H'_{c2}(T_c)$ \cite{Kim2013} where $\phi_0=2.07\times 10^{-7}$ G~cm$^{2}$ is a magnetic flux quantum. We used heat capacity data by Uhlarz et al. \cite{Uhlarz2010}, see supporting material for more details. 
As can be seen from Fig. \ref{fig1}b, the normalized superfluid density $\rho_s$ of Rh$_{17}$S$_{15}$ is very different from expectations of full-gap $s$-wave superconductors (green line).

The penetration depth measurements clearly indicate that a gap with line nodes is realized in~\rhs. Since Knight shift experiments indicate a spin-singlet order parameter~\cite{Koyama2010}, the orbital component of the pair potential must be even parity. A possible gap consistent with the evidence is a sign-changing state in the $A_{1g}$ irrep. Although nodes are not required by the symmetry of the order parameter, ``accidental'' nodes depending on the microscopic details
of the system are possible, as is the case in some pnictide compounds \cite{Yamashita2011}. Specifically, we propose the gap function
\begin{equation}\label{eq:gap}
    \Delta(\hat{\bm{k}}) = \Delta_0 C_r[r+(1-|r|)(\hat{k}_x^4 + \hat{k}_y^4 + \hat{k}_z^4)]
\end{equation}
where $C_r$ is a normalization constant, see Appendix~\ref{subsec:app:extendeds}. This gap has line nodes for $-0.5<r<-0.25$. As shown in Fig.~\ref{fig1}b, the calculated superfluid density~\cite{Prozorov2006} is in quantitative agreement with the experiment for $-0.45\leq r \leq -0.4$. In this range, the gap has circular line nodes centered about the crystal axes as shown in Fig.~\ref{fig4}b (See Appendix~\ref{subsec:app:extendeds} for the evolution of the gap structure with $r$). 
Other nodal states are strongly constrained by the cubic crystal symmetry. 
For example, although the three-dimensional $d$-wave state $\Delta(\hat{\bm k}) = \sqrt{{4}/{15}}\Delta_0(\hat{k}_x^2 - \hat{k}_y^2)$ is consistent with the penetration depth data since it belongs to the two-dimensional $E_g$ irrep, it reduces the symmetry from cubic to tetragonal \cite{Sigrist1991}.
Such a nematic superconducting state is highly exotic, having so far only been observed in the Bi$_2$Se$_3$ family.~\cite{Matano2016} The nematicity is reflected in the superfluid density, which only fits the data for an in-plane field. Although we cannot fully exclude a nematic state in the $E_g$ irrep, the isotropic superfluid density of the $A_{1g}$ state makes the latter a more conservative scenario. Pairing states in other nontrivial irreps are also unlikely since they imply very high angular momentum.  

\begin{figure}
\includegraphics[width=0.95\linewidth]{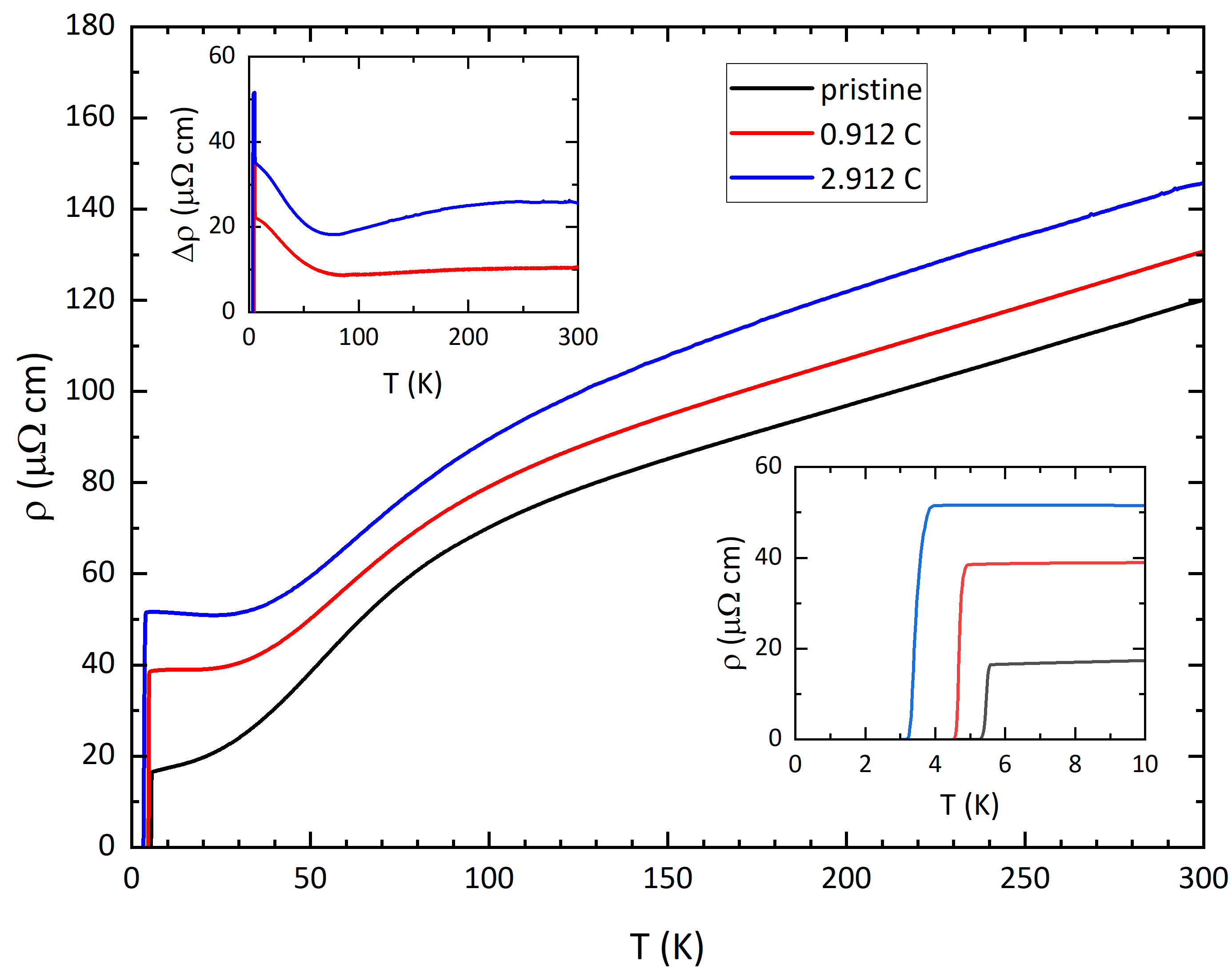}%
\caption{\label{fig2} Effect of electron irradiation on \rhs. (a) Temperature-dependent resistivity of single crystalline sample S2 of Rh$_{17}$S$_{15}$ before irradiation in the pristine state (black line) and after 0.912 C/cm$^2$ (red line) and 2.912 C/cm$^2$ (blue line) electron irradiation. The upper inset shows the temperature-dependent shift of the resistivity curve as a result of irradiation, revealing clear Matthiessen's rule violation at low temperatures, and its validity near room temperature. The lower inset zooms the superconducting transition.
}
\label{resistivity}
\end{figure}

A test for the accidental node scenario can be made by studying the $T_c$ suppression rate with the disorder. For this purpose, the estimation of the dimensionless scattering rate is required, which can be determined from resistivity measurements \cite{Prozorov2014}.
Figure \ref{fig2} shows temperature-dependent resistivity of single crystalline sample S2 of Rh$_{17}$S$_{15}$ before irradiation in the pristine state (black line) and after electron irradiations with doses of 0.912 C/cm$^2$ (red line) and 2.912 C/cm$^2$ (blue line), respectively. The upper inset shows the temperature-dependent shift of the resistivity $\Delta\rho(T)$ as a result of irradiation, clearly indicating the introduction of scattering centers. Matthiessen's rule is valid only above 100 K and 200 K for 0.912 C and 2.912 C, respectively, and it is violated at low temperatures. The lower inset shows a zoom of the superconducting transition, and $T_c$ in the irradiated samples decreased, consistent with the presence of the anisotropic superconducting gap in \rhs~suggested by London penetration depth measurements.

The clear downturn of the $\rho(T)$ curve on cooling below 100~K from roughly $T-$linear at high temperatures is accompanied by sign change of the Hall effect \cite{Naren2008,Daou2016} and the emergence of notable non-linearity of its field dependence. Analysis of the field-dependent resistivity and Hall effect in a multi-band scenario finds at least two groups of carriers with notable differences in properties, acting in parallel. The group of high mobility carriers with mobilities up to 600 cm$^2$/(V sec) dominates transport at low temperatures, while carriers with normal for metals mobilities of order 1 cm$^2$/(V sec) are responsible for transport at high temperatures \cite{Daou2016}. 
This extreme difference in the properties of the carrier makes it hard to determine a single effective scattering rate from resistivity change as usually done \cite{Prozorov2014}.
The difference in mobilities naturally explains a significantly bigger increase of resistivity after irradiation at low temperature \cite{Bass1972,BaRuTanatar}.

Although the shoulder-like feature in Fig.~\ref{fig2} is rather smooth and broad, somewhat sharper features similar to this are observed in three-dimensional materials undergoing charge density wave instability \cite{Naito1982}.
High mobility carriers here may arise from small pockets forming upon Fermi surface reconstruction. Though we do not find any sharp features suggesting phase transition in Rh$_{17}$S$_{15}$, out of an abundance of care we made sure that there are now new diffraction peaks suggesting structural transition down to 5~K (see Appendix).

\begin{figure}
\includegraphics[width=1\linewidth]{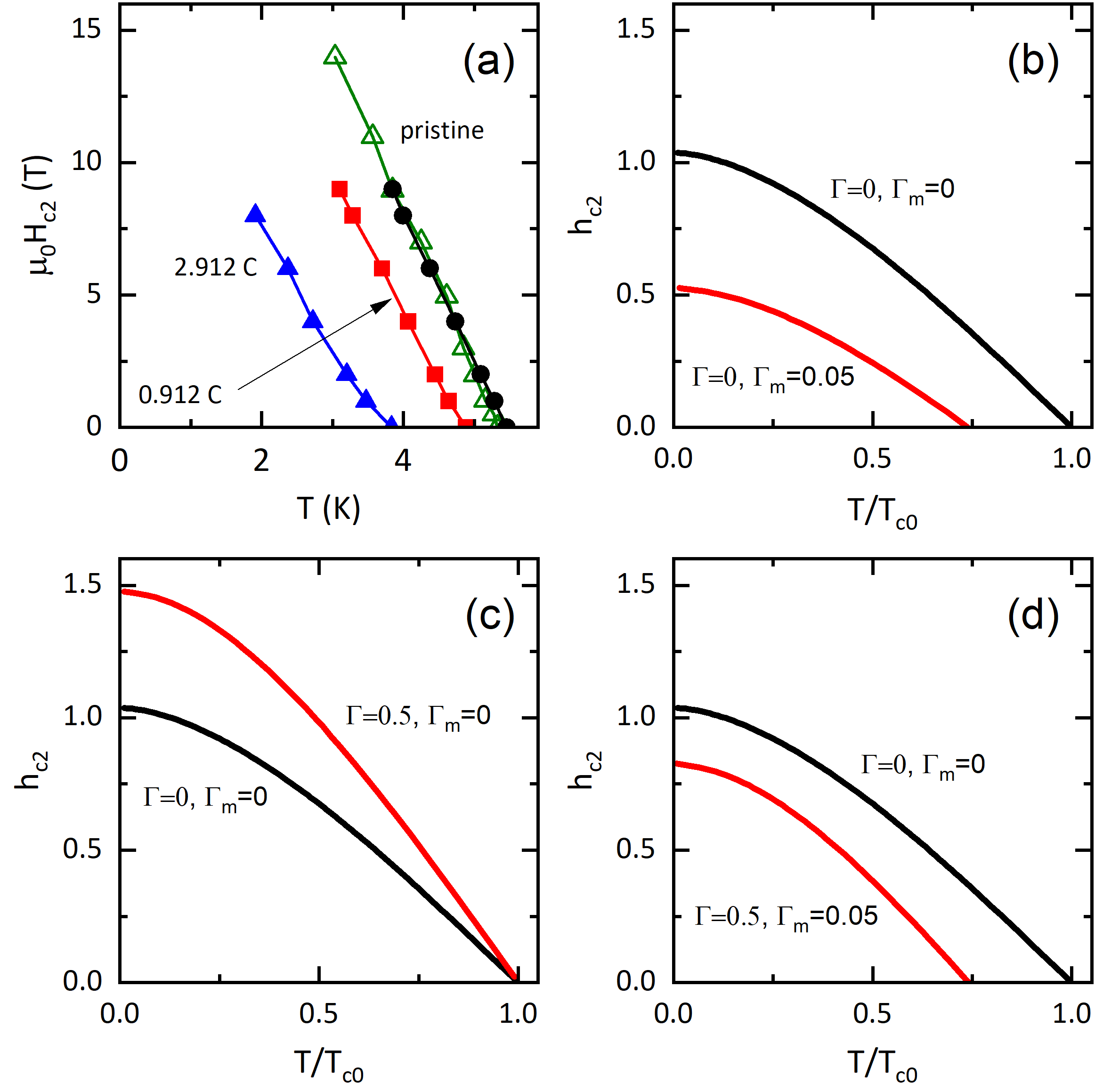}%
\caption{\label{fig3} (a) Temperature-dependent upper critical field in pristine (black curve) and electron-irradiated (red curve) single crystals of Rh$_{17}$S$_{15}$. The onset of resistivity was used as a criterion. The full dataset is presented in the supplementary material. Measurements were taken in magnetic fields parallel to [100] crystallographic direction. The green open symbols show $H_{c2}$ determined from heat capacity measurements in polycrystalline samples \cite{Uhlarz2010}. (b-d) The evolution of $H_{c2}(T)$ with the pairbreaking ($\Gamma_m$) and non-pairbreaking ($\Gamma$) scattering. The corresponding scattering rates are indicated in each panel.
}
\end{figure}

The relatively high upper critical field of Rh$_{17}$S$_{15}$, $H_{c2}(0) \approx $20 T \cite{Naren2011SUST}, suggests that the carriers involved in superconducting pairing should be rather heavy, since $H_{c2} \sim v_F^{-2}$, where $v_F$ is Fermi velocity \cite{Helfand1966,Kogan2012}. In contrast, the London penetration depth and resistivity are dominated by light carriers. To access properties and response to the disorder of heavier carriers in the condensate, we measured the upper critical field, and the results are shown in Fig.~\ref{fig3}a. In a pristine state, $H_{c2}(T)$ determined from the onset of resistivity in our measurements (full symbols) matches perfectly with the entropy-balance definition from the specific heat measurements \cite{Uhlarz2010}. After the irradiation, the $H_{c2}(T)$ curves are shifted to lower temperatures with slightly decreased slope, matching the BCS expectation of $H_{c2}(0) \sim T_{c0}$ \cite{Kogan2012}. The theoretical curves illustrating this case are shown in Fig.~\ref{fig3}b. Here $\Gamma$ is the dimensionless potential (non-pairbreaking) scattering rate and $\Gamma_m$ is a pair-breaking one. If we had a non-pairbreaking situation, the $H_{c2}(T)$ would shift upward, as shown theoretically in Fig. \ref{fig3}c \cite{Kogan2022}. In addition to pure pair-breaking and non-pairbreaking cases, panel Fig.~\ref{fig3}d shows the mixed case when both types of scattering channels are present. While $T_c$ still decreases, the slope of $H_{c2}(T)$ increases inconsistent with our data. We, therefore, conclude that the induced non-magnetic disorder is pairbreaking, compatible with our model gap function.
\begin{figure}
\includegraphics[width=0.8\linewidth]{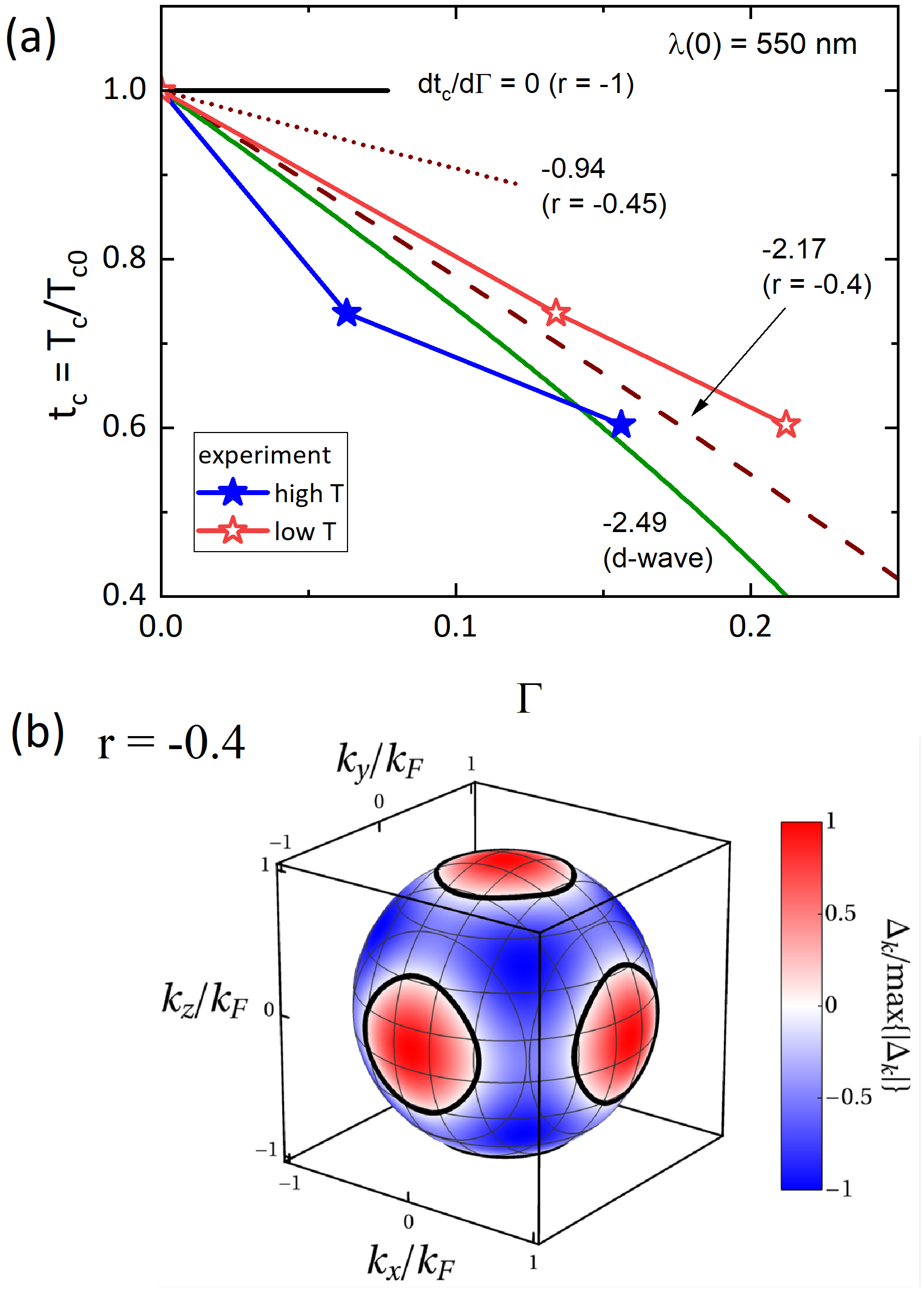}%
\caption{\label{fig4} Pair-breaking effect and superconducting gap in \rhs. (a) $T_c/T_{c0}$ vs. $\Gamma$ for various $r$ in \rhs~compared to the typical gap symmetries. See text for the definitions of $\Gamma$ and $r$. The closed and open star symbols represent the calculated $\Gamma$ with $\Delta\rho$ (see Fig. 2) at room temperature and low temperature, respectively.
(b) Superconducting gap structure in \rhs~with $r=-0.4$.}
\end{figure}

Figure \ref{fig4}a shows the suppression of $T_c$ with scattering for various $r$, introduced in Eq. (2). The scattering rate is represented by a dimensionless parameter $\Gamma$ \cite{Prozorov2014}. The $T_c$ suppression rate increases with decreasing $|r|$. It is notable that $dt_c/d\Gamma$ with $r=-0.4$ is similar to the expectation for $d$-wave. Although the superfluid densities with the choices of $r=-0.4$ and $-0.45$ both yield good agreement with the experiment, their response to the impurity scattering is drastically different. Hence, the experimental determination of the suppression rate is helpful to get insight into the superconducting pairing state in \rhs. However, the direct determination of $\Gamma$ is currently unavailable for multi-band systems. Instead, we test our experimental observation within a single band model by using the relation below \cite{Prozorov2014}. 
\begin{equation}
    \Gamma = \frac{\hbar \Delta\rho_0}{2\pi k_B \mu_0 T_{c0} \lambda^2(0)}
\end{equation}
Here we use $T_{c0}=5.4$ K and $\lambda(0)=550$ nm. Determination of $\rho_0$ is not trivial because of the strong temperature dependence of $\Delta \rho$ below 100 K, and, therefore, we evaluate $\Gamma$ at low temperature just above $T_c$ as well as room temperature, and the resultant initial slope $dt_c/d\Gamma$ are $-2$ and $-4$, respectively. Whereas the finite $T_c$-suppression rates with the non-saturating $\Delta \lambda(T)$ undoubtedly indicate the nodal superconductivity in \rhs, it exhibits a substantial uncertainty depending on $\Delta\rho_0$. We attribute the uncertainty to the multiband nature of \rhs~where an accurate determination of $\Gamma$ is difficult due to different scattering rates for each band. This may result in an overestimate of $\Gamma$ within a single band model.


In conclusion, we discovered a nodal superconducting energy gap in cubic $4d$-electron superconductor \rhs~evidenced by almost linear temperature variation of London penetration depth at low temperatures, and the disorder-response in the superconducting critical temperature, electrical resistivity, and upper critical field. The calculated normalized superfluid density is consistent with an extended $s$-wave gap with ring-shaped accidental nodes. Our results suggest that \rhs~is the first known nodal superconducting gap in a mineral superconductor and add \rhs~to a rare case of nodal superconductor in the $4d$-electron cubic system. The discovery of unconventional superconductivity in \rhs~has important implications for the field of condensed matter physics. It highlights the rich variety of superconducting materials and opens up new possibilities for exploring the interplay between unconventional superconductivity, strong electron-boson coupling, and the role of rhodium in superconducting compounds. Further investigations on \rhs~and related materials may lead to a better understanding of the fundamental principles governing superconductivity and pave the way for developing novel superconducting materials with improved properties and potential applications in various fields.

Finally, we note that naturally occurring single crystals of miassite have interestingly notable content of iron, nickel, platinum, and copper impurities \cite{impurities}, at a level of a few percent. Considering the unconventional character of superconductivity as determined in this study, they are not expected to show any superconductivity in the natural form. Nature knows how to hide its secrets!

\begin{acknowledgments}
PCC acknowledges the Encyclopedia of Minerals (Second Edition), by W. L. Roberts, T. J. Campbell and G. R. Rapp as a continued source of inspiration.
PCC and SLB acknowledge Xiao Lin for having helped to initiate this line of research.
HK is grateful for the useful discussions with Johnpierre Paglione and Daniel Agterberg. Work in Ames was supported by the U.S. Department of Energy (DOE), Office of Science, Basic Energy Sciences, Materials Science and Engineering Division. Ames Laboratory is operated for the U.S. DOE by Iowa State University under contract DE-AC02-07CH11358.
\end{acknowledgments}

\bibliographystyle{apsrev4-2}
\bibliography{Rh17S15}

\newpage

\appendix

\section{Electron irradiation}
The low-temperature 2.5 MeV electron irradiation was performed at the SIRIUS Pelletron facility of the Laboratoire des Solides Irradi\'es (LSI) at the Ecole Polytechnique in Palaiseau, France. 
The acquired irradiation dose is conveniently expressed in C/cm$^2$ and measured directly as a total charge accumulated behind the sample by a Faraday cage. 
Therefore, 1 C/cm$^2$ $\approx$ $6.24 \times 10^{18}$ electrons/cm$^2$. 
In the experiment, the London penetration depth was measured, then the sample was irradiated, and the cycle was repeated. 
The irradiation was carried out with the sample immersed in liquid hydrogen at about 20 K. 
Low-temperature irradiation is needed to slow down the recombination and migration of defects. Upon warming up to room temperature, a quasi-equilibrium population of atomic vacancies remains due to a substantial difference in the migration barriers between vacancies and interstitials. 
An example of such incremental irradiation/measurement sequence showing the resistivity change measured in-situ, as well as the annealing after warming up, is given elsewhere \cite{Prozorov2014}. 
In the present case, the sample was dispatched between the lab and the irradiation facility for the measurements and irradiation, and then the sequence was repeated. Further information on the physics of electron irradiation can be found elsewhere \cite{Damask1963, Thompson1969}.

\section{Electrical transport measurement}
Four-probe electrical resistivity measurements were performed in {\it Quantum Design} PPMS on three samples S2, S3, and S4. Contacts to the samples were soldered with tin \cite{Tanatar2010} and had resistance in m$\Omega$ range. Electrical resistivity at room temperature was determined as 130$\pm$20 $\mu \Omega$ cm, based on the average of three samples. 
The same samples with contacts were measured before and after electron irradiation, excluding the uncertainty of the geometric factor determination.

\section{Tunnel diode resonator technique}

The temperature variation of London penetration depth was measured in a $^3$He cryostat and a dilution refrigerator (DR) by using a TDR technique \cite{Prozorov2011}. 
A small sample typically  $<$ 0.8 mm in the longest dimension was mounted on a sapphire rod with a diameter of 1 mm and inserted into a 2 mm inner diameter copper coil that produces rf excitation field with empty-resonator frequency of 14.5 MHz and 16.6 MHz for $^3$He-TDR and DR-TDR, respectively, with amplitude $H_{ac} \approx 20$ mOe. The shift of the resonant frequency (in cgs units), $\Delta f(T)=-G4\pi\chi(T)$, where $\chi(T)$ is the differential magnetic susceptibility, $G=f_0V_s/2V_c(1-N)$ is a constant, $N$ is the demagnetization factor, $V_s$ is the sample volume and $V_c$ is the coil volume. The constant $G$ was determined from the full frequency change by physically pulling the sample out of the coil. With the characteristic sample size, $R$, $4\pi\chi=(\lambda/R)\tanh (R/\lambda)-1$, from which $\Delta \lambda$ can be obtained \cite{Prozorov2006,Prozorov2011}.

\section{X-ray diffractometry}

\begin{figure}[t]
\includegraphics[width=1\linewidth]{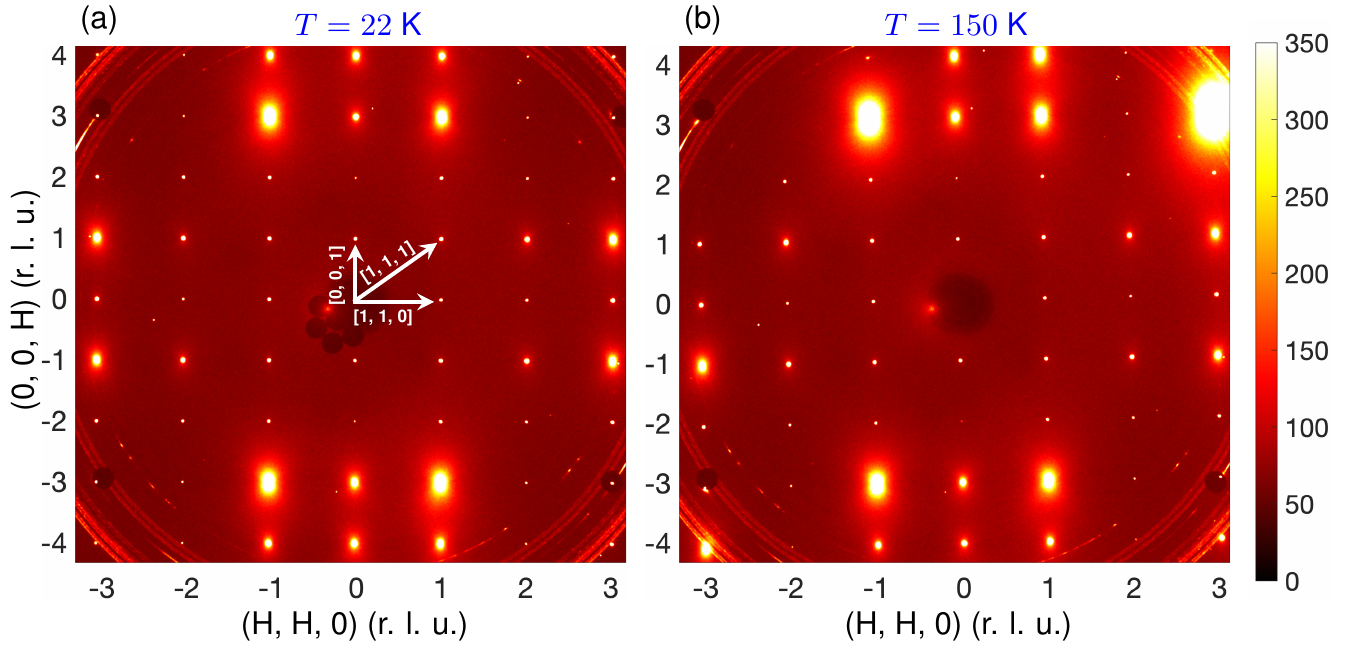}%
\caption{Results of high-energy X-ray diffraction measurements showing extended regions of (H,H,0)-
(0,0,H) reciprocal planes at two temperatures 22 (left) and 150 K (right). Intensities in each panel are color coded to a 
linear scale.  Small circular black
circles at f3, 3, 3g Bragg peaks and around the beam center (in panel \a") are from the lead masks which were
used to mask the intense Bragg peaks to avoid the oversaturation of the detector pixels. Furthermore,
the polycrystalline rings are from the Be domes. In panel b, the large intensity at (3, 3, 3) peak is due to
the lead mask not being at the correct position. Also, the higher intensity for (-1, -1, 3) Bragg peak in
comparison to (-1 -1 -3) and missing (-2, -2, 3) peak are due to the crystal being slightly misaligned at 150 K. The strong regular lattice peaks were partially blocked/saturated to enable resolution in 10$^{-4}$ to 10$^{-5}$ range. No additional superstructure peaks corresponding to CDW were observed at low temperatures within this resolution.}
\label{xray}
\end{figure}

The high-energy X-ray diffraction measurements were performed at station 6-ID-D at the Advanced Photon Source, Argonne National Laboratory. The use of X-rays with an energy of 100 keV minimizes sample absorption and allows to probe of the entire bulk of the sample using an incident beam with a size of 0.5$\times$0.5 mm$^2$, over-illuminating the sample. The samples were held on Kapton tape in a Helium closed-cycle refrigerator and Helium exchange gas was used. Extended regions of selected reciprocal lattice planes were recorded by a MAR345 image plate system positioned 1585 mm behind the sample as the sample was rocked through two independent angles up to $\pm 2.0$\degree~about axes perpendicular to the incident beam.

\section{Transport properties}
The cross-over maximum in $\rho (T)$ at around 100~K may be caused by the contribution of two types of carriers with very different properties, as usually discussed for materials with Matthiessen's rule violation \cite{Bass1972,Sondheimer1947,Allen1980} in "parallel resistor" model. The crossover in resistivity is accompanied by the sign change of the Hall effect \cite{Naren2008,Daou2016}. The large magnitude of the Hall constant at low temperatures suggests that low carrier density electrons dominate transport near $T_c$ \cite{Naren2008}. Estimate from single band model, $R_H=1/(ne)$, the carrier density $n\approx 3\times10^{20}$ cm$^{-3}$ provides an upper bound \cite{Naren2008}. Use of the two-carrier type analysis of the field and temperature-dependent Hall effect and magnetoresistance suggest that electron carrier density is in fact notably lower, $\sim$1$\times$10$^{19}$ cm$^{-3}$ and the mobility of these carriers is quite high, in 600 cm$^2$V$^{-1}$s$^{-1}$ range, indicating small effective mass of the carriers \cite{Daou2016}. These high mobility electronic carriers are clearly inconsistent with rather high specific heat Sommerfeld $\gamma$, a high value of heat capacity jump at $T_c$, and the high upper critical fields. This may be suggesting that holes, dominating the Hall effect at high temperatures, should be notably heavier. The two-band character of the transport clearly reveals itself in response to electron irradiation. The contribution of small carrier density high mobility carriers suffers two times stronger from the introduction of disorder \cite{TanatarJPCM}, giving an increase in the difference plot on cooling. The two-band charge transport needs to be accounted for properly for the estimation of the scattering rate, introduced by irradiation. We adopted resistivity variation at low and high temperatures as an uncertainty of the scattering rate determination. 

\section{Rutgers relation}

The heat capacity jump is $\Delta C/T=0.24$ J/mol K$^{2}$ \cite{Uhlarz2010,Naren2008}. Note we used the lattice parameter of 9.911 \AA \cite{Geller1962} for conversion of the heat capacity jump. The slope is determined to be $H'_{c2}(T_c) = -6.78$ T/K by fitting data points between $0.87T_c$ and $T_c$ \cite{Uhlarz2010}. Choice of $\lambda(0)=550$ nm gives good agreement between experiment and the Rutgers' relation where $\rho_s'(1)=-1.11$.

\section{Upper critical field determination}

\begin{figure}
\includegraphics[width=0.8\linewidth]{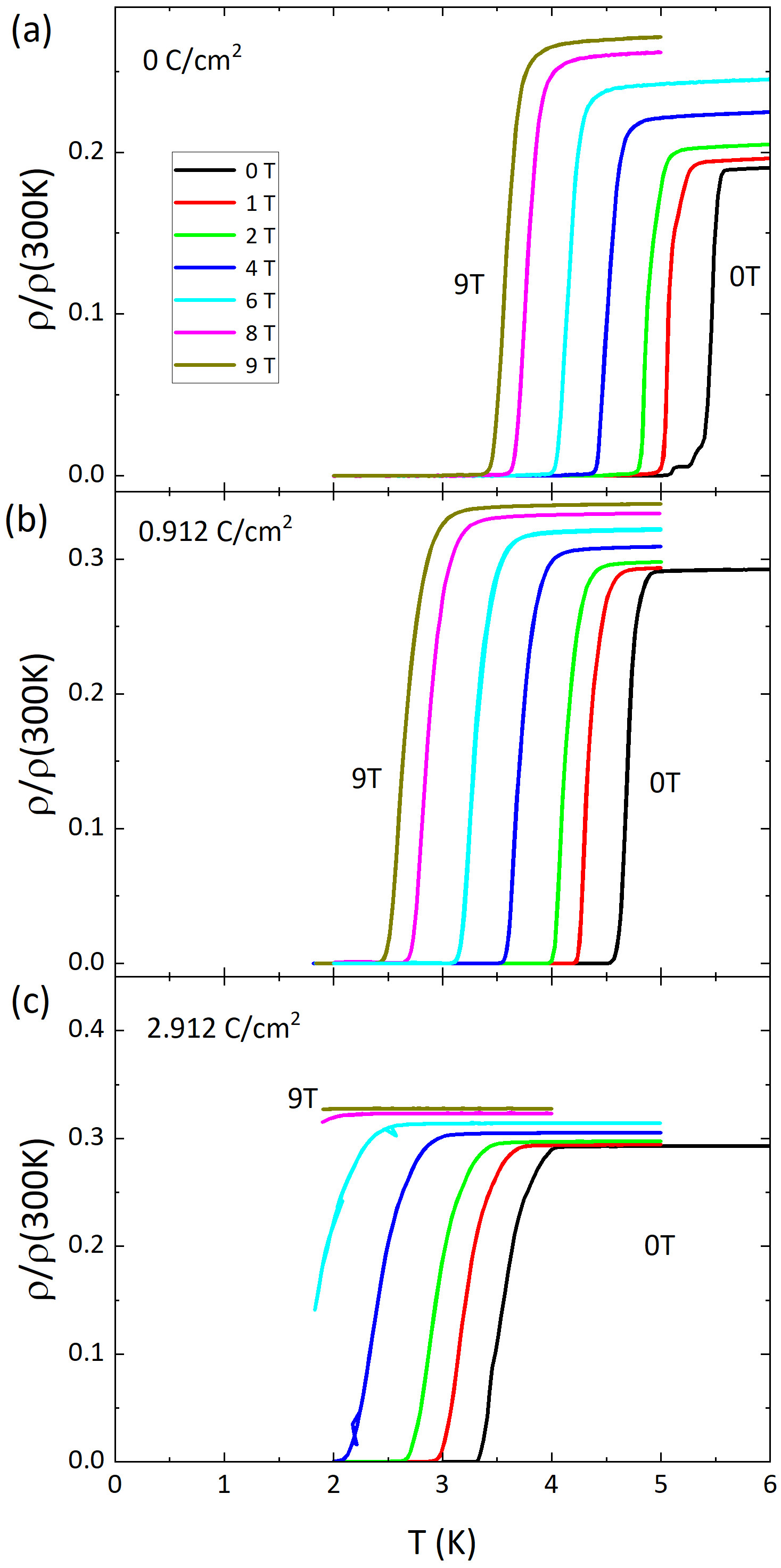}%
\caption{\label{sfig2} Temperature dependent resistivity of pristine (a) 0 C/cm$^2$, and electron irradiated samples (b) 0.912 C/cm$^2$ and (c) 2.912 C/cm$^2$ of Rh$_{17}$S$_{15}$ in magnetic fields 0T, 1 T, 2 T, 4T, 6T, 8 T and 9 T (right to left). The magnetic field was applied transverse to the current and along [100] crystallographic direction. }
\end{figure}

The upper critical fields were determined from resistive measurements. 
In Fig.~\ref{sfig2} we show temperature-dependent resistivity measured in magnetic fields applied along [100] crystallographic direction transverse to the electrical current. The top panel shows results for the pristine sample, bottom panel for the sample with 0.912 C/cm$^2$ irradiation. Resistive transitions remain sharp with the application of magnetic fields and do not show significant broadening. 
We used the onset criterion for $T_c$ determination.
These $H_{c2}$ determinations are in good agreement with the upper critical fields determined from heat capacity measurements in which equal entropy construction was used for the determination of both $T_c$ and $H_{c2}(T)$ \cite{Uhlarz2010}. The resistivity onset curve practically coincides with the heat capacity curve, see Fig.~\ref{fig3}a.

\section{Superconducting state}


Having established the angular dependence of the order parameter, $\Omega(\varphi,\theta)$, we can now solve the Eilenberger self-consistency equation for the temperature-dependent part, $\Psi(t)$, to obtain the superconducting gap function $\Delta=\Psi \Omega$. Details of the calculations of this and other thermodynamic quantities are described in detail elsewhere Ref.\cite{Kogan2009b,Prozorov2011}. For comparison, the application of this approach to MgB$_2$ superconductor is shown in Ref.\cite{Kim2019}.

\subsection{Extended $s$-wave pairing state}\label{subsec:app:extendeds}

\begin{figure}[b]
\includegraphics[width=0.8\linewidth]{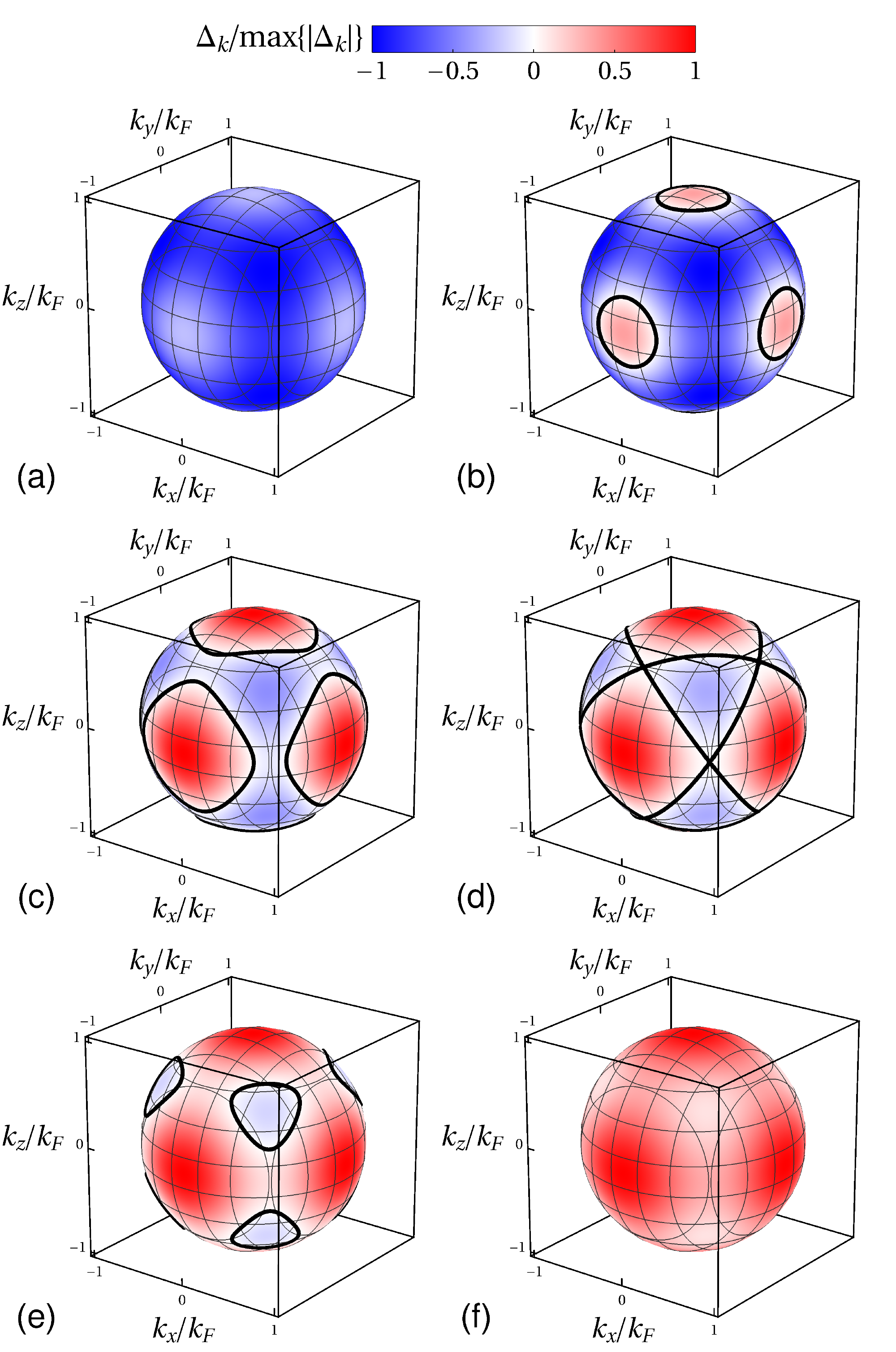}%
\caption{Extended $s$-wave gaps for (a) $r=-0.55$, (b) $r=-0.45$, (c) $r=-0.35$, (d) $r=r_\text{crit}=-\frac{1}{3}$, (e) $r=-0.3$, and (f) $r=-0.2$.}
\label{swavesupp}
\end{figure}

The  angular-dependent part of the order parameter for the extended $s$-wave gap proposed in the paper is

\begin{eqnarray} \label{eq:extswave}
    \Omega(\theta,\phi) & = & C_r\left[r  + (1-|r|)\left( \cos^4\theta  \right. \right. \notag \\
    && \left.  \left. +\sin^4\theta[\sin^4\varphi + \cos^4\varphi]\right)\right] 
\end{eqnarray}
where the normalization constant
\begin{equation}
    C_r = \sqrt{\frac{105}{41 + 126 r + 146 r^2 - 2 (41 + 63 r)|r|}}
\end{equation}
ensures that $\int|\Omega(\theta,\phi)|^2 d\Omega = 4\pi$. The key evolution of $\Omega(\theta,\phi)$ with $r$ is shown in fig.~\ref{swavesupp}, At $r=-1$, the form factor gives isotropic $s$-wave pairing. Increasing $r$, gap minima develop along the crystal axes [panel (a)]. At $r=-0.5$ this deepens to a quadratic point node; further increasing $r$ this point node develops into a circular line node centered around the crystal axes [panels (b) and (c)]. These grow with $r$ until they touch at the critical value of $r_{\text{crit}}=\frac{1}{3}$ [panel (d)]. Further increasing $r$ the line nodes recombine into eight circular line nodes about the cubic body diagonals [panel (e)]. Increasing $r$ beyond $-0.25$ these nodes collapse down to quadratic points, and then eventually give rise to an anisotropic fully-gapped state with gap minima along the cubic body diagonals [panel (f)].

\subsection{Thermodynamic quantities}

\begin{figure}
\includegraphics[width=0.8\linewidth]{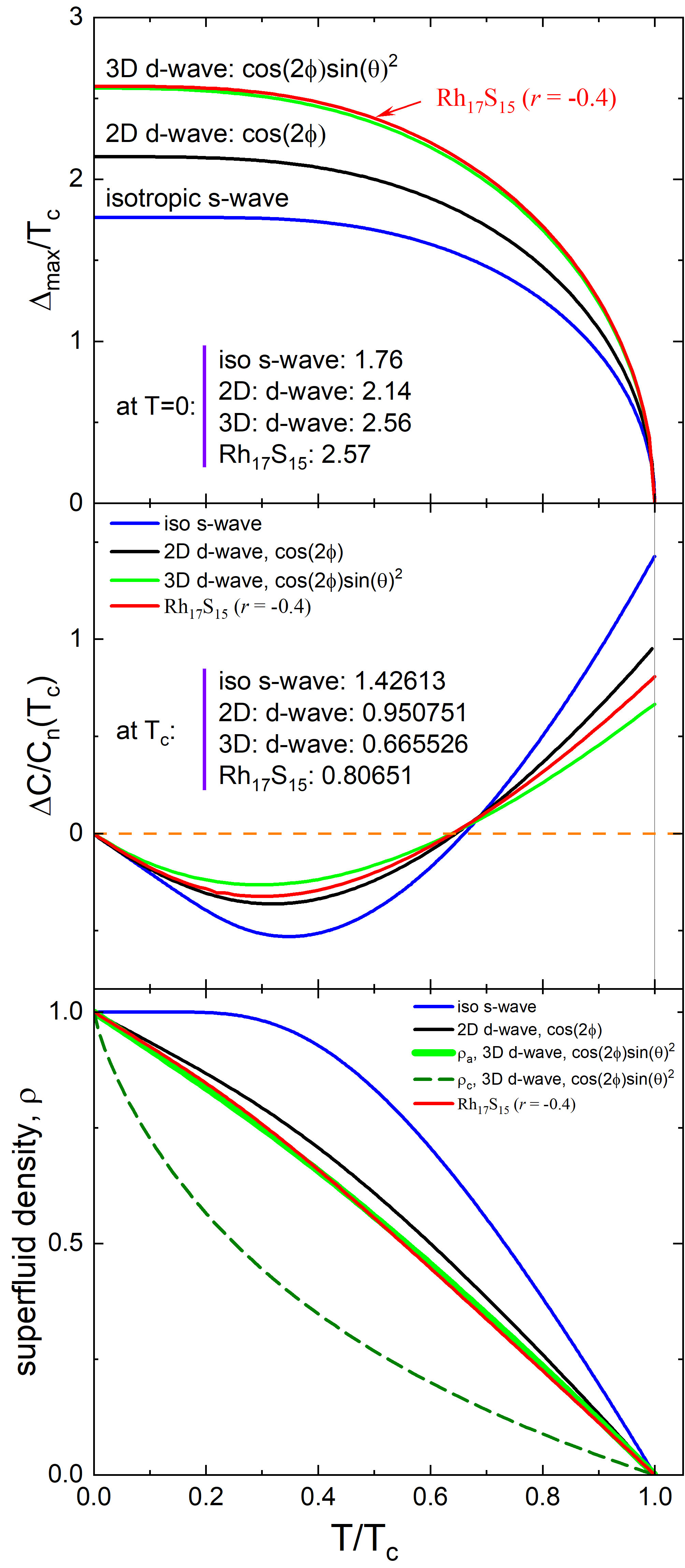}%
\caption{\label{sfig4} Temperature evolution of (a) superconducting gap, (b) heat capacity, and (c) superfluid density in \rhs~ (red solid lines), compared to isotropic $s$-wave and  2D and 3D $d$-wave symmetries.}
\end{figure}
Figure \ref{sfig4} summarizes the temperature dependence of the calculated superconducting gap, heat capacity, and superfluid density for our order parameter with $r=-0.45$, shown by the red solid curves. For comparison, other usual cases are shown on all frames. Notice the difference between 2D s-wave, $\sim \cos({2\varphi})$, and 3D $s$-wave, $\sim \cos({2\varphi})\sin^2(\theta)$. Interestingly, panel (a) shows that the maximum gap value for our order parameter is less than that in typical $s$-and $d$-wave gap symmetries.  
Unfortunately, we are not aware of the direct measurements of the gap \rhs, and the maximum gap value is unknown. On the other hand, the gap estimated from thermodynamic measurements only provides the Fermi surface average.

In panel (b) we plot the temperature dependence of the predicted deviation of the specific heat from the normal state, i.e. $\Delta C/C_n = (C_s - C_n)/C_n$, with $C_n=\gamma T$. For all nodal gaps the predicted heat capacity jump at $T_c$ [$\Delta C(T_c)$] is smaller than the BCS weak-coupling value of 1.43, since the condensation energy is maximal for an isotropic $s$-wave gap. In contrast, the reported experimental heat-capacity jump is nearly 2 \cite{Naren2008}.
High ratios of $\Delta C/C_n > 2$ have been found in heavy fermion compounds, namely CeCoIn$_5$, CeRhIn$_5$, U$_6$Fe, UBe$_{13}$, PrOs$_4$Sb$_{12}$, NpPd$_5$Al$_2$ (see Ref. \cite{White2015} for review). The origin of these high heat capacity jumps is not well understood, but exceeding the weak-coupling value is typically associated with strong-coupling superconductivity; performing a strong-coupling calculation for~\rhs is far beyond the scope of the current work.
Naren {\it et al.} suggest the Sommerfeld coefficient $\gamma\approx 100$ mJ/mol K$^2$ in \rhs~is larger than that of conventional metals possibly due to a narrow Rh-$d$ band 
{consistent with the observed high $H_{c2}(0)$ indicating heavy effective mass}, 
and the shoulder-like feature around 100 K in the resistivity is associated with the formation of the narrow band \cite{Naren2008}. However, the response of the resistivity to disorder revealed the disobeyed Matthiessen rule below 100 K, suggesting the absence of activation that is expected for narrow band contribution.

\begin{figure}
\includegraphics[width=1\linewidth]{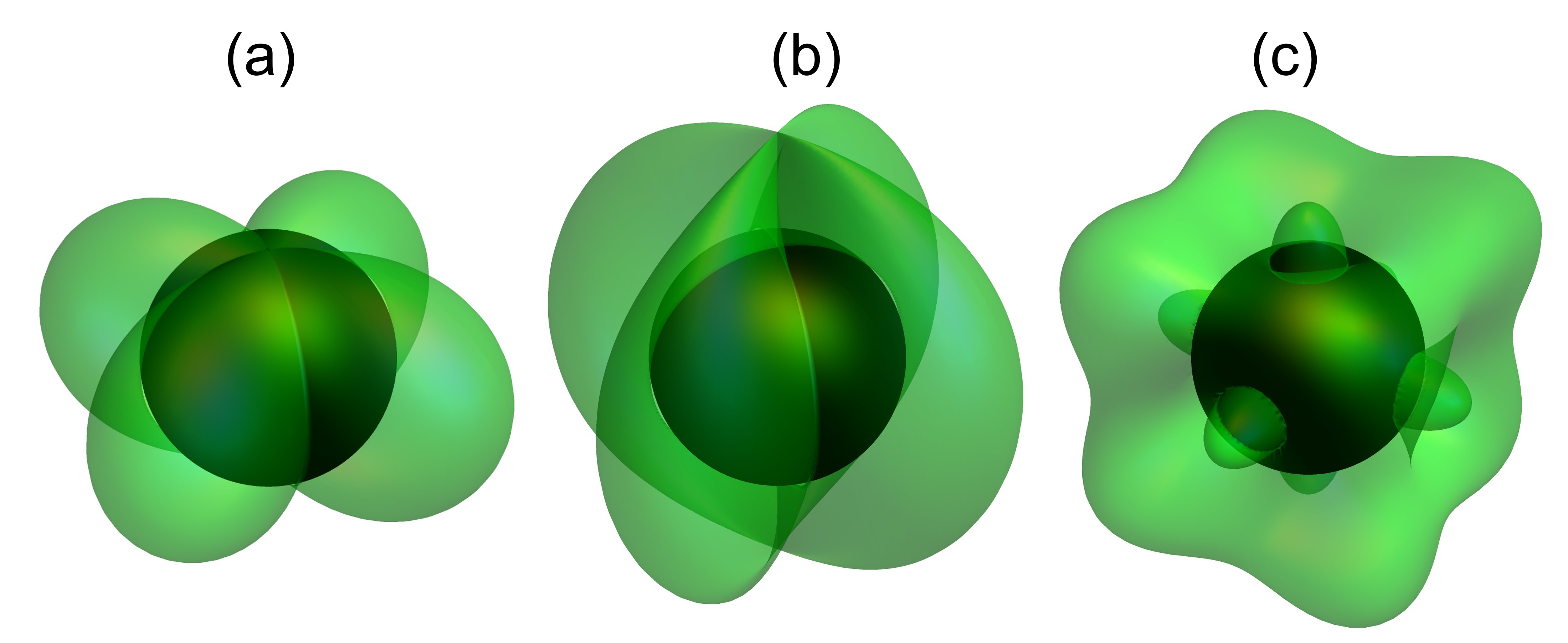}%
\caption{\label{sfig5} Possible superconducting gap on a spherical Fermi surface (dark green) relevant for Rh$_{17}$S$_{15}$. (a) 3D $d$-wave, $\cos(2\phi)\sin^2\theta$. (b) 2D $d$-wave, $\cos(2\phi)$. Clearly, the 2D case is impossible because it results, mathematically, in a 4-fold degenerate vertical line node not residing on any Fermi surface. Panel (c) is for the extended $s$-wave with accidental nodes (see Eq. \ref{eq:extswave}).}
\end{figure}

In panel (c), we compare the calculated superfluid density in various gap symmetries. The superfluid density \rhs~is remarkably well reproduced by an in-plane component of a 3D $d$-wave. We note that there are no free parameters here. However, since this gap belongs to the two-dimensional $E_g$ irrep, it lowers the symmetry of the electronic dispersion from cubic to tetragonal, implying a nematic superconducting state. This is a highly exotic scenario, as nematic superconductivity has so far only been observed in the Bi$_2$Se$_3$ family~\cite{Matano2016}. The nematicity is reflected in the superfluid density which shows stark differences between the in-plane and $c$-axis directions.

\end{document}